\documentclass[letterpaper,10pt]{article}

\usepackage{amsmath}
\usepackage{amssymb}
\usepackage{psfrag}
\usepackage[dvips]{graphicx}
\usepackage{epsfig}

\newcommand{\dd}{\textrm{d}}
\newcommand{\im}{{\mathbb{I}}{\mathrm{m}}}

\author{
A. L\'opez-Ortega\thanks{alopezo@ipn.mx} \\
Centro de Investigaci\'on en Ciencia Aplicada y Tecnolog\'{\i}a Avanzada. \\ 
Unidad Legaria. Instituto Polit\'ecnico Nacional. \\
Calzada Legaria \# 694. Colonia Irrigaci\'on. Delegaci\'on Miguel Hidalgo. \\
M\'exico, D.\ F., M\'exico. \\
C.\ P.\  11500  
}

\title{On the quasinormal modes of the de Sitter spacetime}

\begin{document}

\maketitle

\begin{abstract}

Modifying a method by Horowitz and Hubeny for asymptotically anti-de Sitter black holes, we establish the classical stability of the quasinormal modes of the de Sitter spacetime. Furthermore using a straightforward method we calculate the de Sitter quasinormal frequencies of the gravitational perturbations and discuss some properties of the radial functions of these quasinormal modes.



\end{abstract}

\section{Introduction}
\label{s: Introduction}

The de Sitter spacetime is an extensively studied solution of the gravitational field equations. It is a maximally symmetric spacetime and due to its simplicity it is the appropriate background to explore and test several ideas and methods \cite{Spradlin:2001pw}, \cite{Kim:2002uz}. 

If a spacetime with an event or cosmological horizon is perturbed, it is well known that the perturbation oscillates with a set of decaying modes, usually called quasinormal modes (QNM), whose complex frequencies are determined by the physical parameters of the background. They have been calculated for several spacetimes (see the reviews \cite{Kokkotas:1999bd}-\cite{Konoplya:2011qq}) and they are useful in the study of several physical phenomena \cite{Kokkotas:1999bd}-\cite{Konoplya:2011qq}. 

The quasinormal frequencies (QNF) of the de Sitter spacetime have been determined in several references \cite{Du:2004jt}--\cite{LopezOrtega:2007sr}. We know the QNF of classical fields propagating in lower \cite{Du:2004jt}--\cite{Zelnikov:2008rg} and higher \cite{Natario:2004jd}--\cite{LopezOrtega:2007sr} dimensional de Sitter spacetimes. Moreover, the recent reviews \cite{Berti:2009kk}, \cite{Konoplya:2011qq} examine some of these results. For $D$-dimensional ($D \geq 4$) de Sitter spacetimes, in Ref.\ \cite{Natario:2004jd} it is asserted that for $D$ even, the QNF of the gravitational perturbations are not well defined, and we notice that this affirmation is reproduced in the review \cite{Berti:2009kk} (see also Appendix A of \cite{Choudhury:2003wd} for a similar conclusion for the massless Klein-Gordon field in the four-dimensional de Sitter spacetime). Nevertheless in Ref.\  \cite{Lopez-Ortega:2006my} it is asserted that for even and odd dimensional de Sitter spacetimes the QNF of massless fields are well defined (a similar affirmation appears in the review \cite{Konoplya:2011qq}).

The values of the de Sitter QNF have found several applications, for example, they have been used in the calculation of the Green functions that are necessary to discuss some aspects of static patch holography \cite{Anninos:2011af}, in the computation of the one-loop thermal partition function for the scalar field \cite{Denef:2009kn}, and in the determination of the entropy spectrum for the de Sitter horizon \cite{LopezOrtega:2009ww}, \cite{LopezOrtega:2011dw}. Thus we must state clearly the existence and properties of the de Sitter QNM.

Here using a different method we prove the classical stability of the de Sitter QNM. This method is motivated by a procedure used to show the stability of the QNM in asymptotically anti-de Sitter spacetimes \cite{Horowitz:1999jd}. Moreover, to solve the mentioned issue on the existence of the de Sitter QNF in even dimensions and employing a straightforward procedure we determine the de Sitter QNF of the gravitational perturbations. Finally we discuss some properties of the radial functions for the gravitational de Sitter QNM.

We organize this paper as follows. In Sect.\ \ref{s: stability} we establish the classical stability of the de Sitter QNM. To achieve this goal we modify a method by Horowitz and Hubeny that is suitable for asymptotically anti-de Sitter black holes \cite{Horowitz:1999jd}. To prove the classical stability of the QNM for the Klein-Gordon and Dirac fields we find that it is necessary to elaborate on the method by Horowitz and Hubeny. In Sect.\ \ref{s: gravitational QNM} we calculate the QNF of the gravitational perturbations propagating in the $D$-dimensional de Sitter spacetime. We compute these with a simpler method than the procedure used in the previous references and discuss some properties of the radial functions for the de Sitter QNM. In  Sect.\ \ref{s: discussion} we enumerate our results. Finally, in Appendix we analyze the absence of QNF for the Klein-Gordon field with special values of the mass and the azimuthal number that propagates in four-dimensional de Sitter spacetime.

\section{Stability of the de Sitter QNM}
\label{s: stability}

In what follows we consider the $D$-dimensional ($D \geq 4$) spherically symmetric de Sitter spacetime \cite{Spradlin:2001pw}, \cite{Kim:2002uz}
\begin{equation} \label{e: metric de Sitter spacetime}
 \dd s^2 = (1-r^2)\dd t^2 - (1-r^2)^{-1} \dd r^2 - r^2 \dd \Omega^2_{D-2},
\end{equation}  
where $\dd \Omega^2_{D-2}$ is the round metric on the $(D-2)$-sphere. As is well known, this background has a cosmological horizon. In the units chosen in the metric (\ref{e: metric de Sitter spacetime}) the horizon is located at $r=1$. We recall that in a spherically symmetric background the equations of motion for the perturbations simplify to Schr\"odinger type equations \cite{Kokkotas:1999bd}--\cite{Konoplya:2011qq}, \cite{Chandrasekhar book}, \cite{LopezOrtega:2009qc}
\begin{equation} \label{e: Schrodinger equation}
 \frac{\dd^2 U}{\dd r_*^2} + \omega^2 U -V U = 0,
\end{equation} 
where $r_*$ is the tortoise coordinate or is related to it, $V$ is an effective potential that depends on the perturbation, and $U$ is related to the radial function of the perturbation. For the discussion that follows, in Table \ref{tab: effective potentials} we enumerate the effective potentials of several fields propagating in de Sitter spacetime \cite{Natario:2004jd}--\cite{LopezOrtega:2007sr}.


\begin{table}
\centering
\caption{Effective potentials in $D$-dimensional de Sitter spacetime. We denote the mass of the field by $\mu$, $q$ is a nonnegative integer, $f=1-r^2$, and $\lambda =\pm (q + (D-2)/2)$ \cite{Natario:2004jd}--\cite{LopezOrtega:2007sr}.}
\begin{tabular}[htp]{ll}
\hline
Field & Effective potential  \\ \hline
Klein-Gordon & $V_{KG} =  f\left( \frac{(D-2)(D-4)}{4 r^2} f -(D-2) +\frac{q(q+D-3)}{r^2} + \mu^2  \right) $ \\ 
Electromagnetic scalar & $V_{EMS}= f \left( \frac{q(q+D-3)}{r^2} +  \frac{(D-2)(D-4)}{4 r^2} - \frac{(D-6)(D-4)}{4} \right)$  \\ 
Electromagnetic vector & $V_{EMV}= f \left( \frac{q(q+D-3) -1}{r^2} +  \frac{(D-2)^2 - 2(D-2) + 4}{4 r^2} - \frac{(D-2)(D-4)}{4} \right)$ \\ 
Gravitational tensor & $V_{GPT}= f \left( \frac{(2 q + D -4)(2 q + D -2)}{4 r^2} - \frac{D(D-2)}{4} \right)$  \\ 
Gravitational vector & $V_{GPV}= f \left( \frac{(2 q + D -4)(2 q + D -2)}{4 r^2} - \frac{(D-4)(D-2)}{4} \right)$  \\ 
Gravitational scalar & $V_{GPS}= f \left( \frac{(2 q + D -4)(2 q + D -2)}{4 r^2} - \frac{(D-4)(D-6)}{4} \right)$  \\ 
Dirac &  $V_{\pm} = \frac{f(\lambda^2 + \mu^2 r^2)^3}{r^2 (\lambda^2 + \mu^2 r^2 + \lambda \mu f / (2 \omega) )^2} \pm \frac{ f^{3/2} (\lambda^2 + \mu^2 r^2)^{5/2}}{\lambda^2 + \mu^2 r^2 + \lambda \mu f / (2 \omega)} $ \\
 & \,\,\,\,\,\,\,\,$\times \left[ \frac{ \frac{1}{2 f} \frac{ \dd f}{\dd r} + \frac{3 \mu^2 r}{\lambda^2 + \mu^2 r^2} }{r(\lambda^2 + \mu^2 r^2 + \lambda \mu f / (2 \omega) )} - \frac{ \lambda^2 + 3 \mu^2 r^2 + \frac{\lambda \mu r}{2 \omega} \frac{\dd f}{\dd r} + \frac{\lambda \mu f}{2 \omega} }{ r^2 (\lambda^2 + \mu^2 r^2 + \lambda \mu f / (2 \omega) )^2 } \right]   $\\ \hline
\end{tabular}
\label{tab: effective potentials}
\end{table}

Here we define the de Sitter QNM as the solutions to Eq.\ (\ref{e: Schrodinger equation}) that satisfy: 

(a) They are purely outgoing near the de Sitter horizon.

(b) They go to zero at $r=0$.  

Notice that the boundary condition at the origin (b) is slightly different from that of \cite{Natario:2004jd}, \cite{Lopez-Ortega:2006my}. In those references it is imposed the boundary condition \textit{``the radial function is regular at $r=0$''}, in particular the radial function can be equal to a constant different from zero. Here we are imposing the boundary conditions (a) and (b) on the function $U$ that appears in the Schr\"odinger type equation (\ref{e: Schrodinger equation}). 

The motivation for the boundary conditions (a) and (b) is as follows. Near the cosmological horizon $(r=1)$ the effective potentials of Table \ref{tab: effective potentials} behave as
\begin{equation} \label{e: limit potential near horizon}
 \lim_{r \to 1} V = 0 ,
\end{equation}  
hence near the cosmological horizon the solution of Eq.\ (\ref{e: Schrodinger equation}) is
\begin{equation} \label{e: solution near the horizon}
 U = C_1 \textrm{e}^{-i \omega r_*} + C_2 \textrm{e}^{ i \omega r_*} ,
\end{equation} 
where $C_1$ and $C_2$ are constants. Thus near the horizon the solution (\ref{e: solution near the horizon}) represents ingoing and outgoing waves. For the de Sitter QNM we must take the purely outgoing wave.  

Near $r=0$, almost all the effective potentials of Table \ref{tab: effective potentials} behave in the form (see below for some exceptions)
\begin{equation} \label{e: potential near zero}
 \lim_{r \to 0} V \to + \infty,
\end{equation} 
and therefore for the potentials that satisfy this condition it is reasonable to impose the boundary condition $U(0)=0$.

\begin{figure}[tbp]
\begin{center}
\includegraphics[clip]{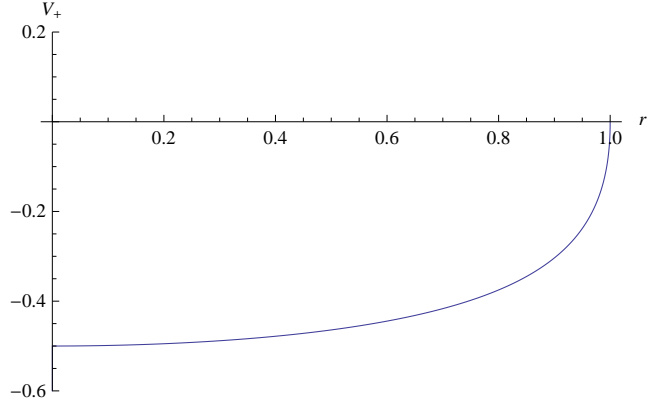}
\caption{For the four-dimensional de Sitter spacetime we draw the effective potential $V_+$ of the massless Dirac field with $\lambda = 1$.}
\label{fig: Dirac vplus lambda 1}
\end{center}
\end{figure}

One exception to the behavior near $r=0$ given in the formula (\ref{e: potential near zero}) is that of the effective potential $V_+$ for the Dirac field with $\lambda =  1$ moving in four-dimensional de Sitter spacetime, since in this example we get
\begin{equation} \label{e: limit zero V+}
 \lim_{r \to 0} V_+ =  \frac{ \frac{3 \mu^2}{2} - \frac{1}{2} +\frac{  7 \mu^3 +  3 \mu}{4 \omega }  }{\left( 1 + \frac{ \mu}{2 \omega} \right)^3} ,
\end{equation} 
and for $\mu = 0$ we obtain
\begin{equation} \label{e: limit zero V+ massless}
 \lim_{r \to 0} V_+ = -\frac{1}{2} .
\end{equation} 
For the effective potential $V_+$ both limits (\ref{e: limit zero V+}) and (\ref{e: limit zero V+ massless}) are finite at $r=0$.  In Fig.\ \ref{fig: Dirac vplus lambda 1}, for the massless Dirac field with $\lambda = 1$, we plot the effective potential $V_+$ in four-dimensional de Sitter spacetime. We see that for these values of $\mu$, $\lambda$, and $D$ the effective potential $V_+$ is negative and that near $r=0$ and $r=1$ its behavior  is consistent with the analytical limits (\ref{e: limit potential near horizon}) and (\ref{e: limit zero V+ massless}).


\begin{figure}[tbp]
\begin{center}
\includegraphics[clip]{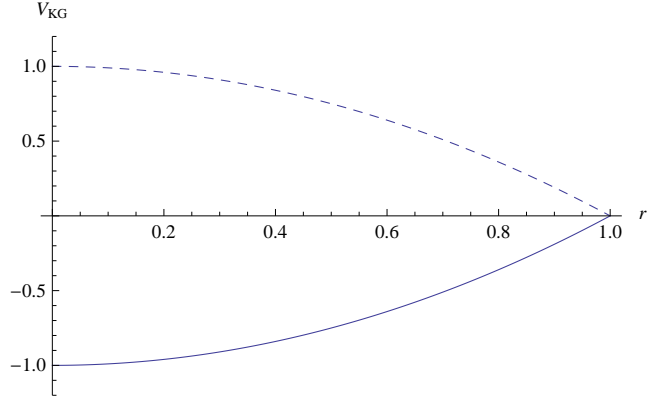}
\caption{For the four-dimensional de Sitter spacetime we plot the effective potentials $V_{KG}$ of the Klein Gordon field  with $q=0$, $\mu^2 = 3$ (dashed line), and $\mu^2 = 1$ (continuous line).}
\label{fig: Klein Gordon D 4 q 0}
\end{center}
\end{figure}

Furthermore in four-dimensional de Sitter spacetime, for the massive Klein-Gordon field with $q=0$ we find that its effective potential becomes (see Table \ref{tab: effective potentials})
\begin{equation} \label{e: effective potential Klein Gordon D4}
 V_{KG} = (\mu^2 - 2) (1-r^2) ,
\end{equation} 
and we point out that this potential is finite at $r=0$. Also for $\mu^2 < 2$ the effective potential (\ref{e: effective potential Klein Gordon D4}) is negative (see Fig.\ \ref{fig: Klein Gordon D 4 q 0}). As far as we have analyzed there is no another effective potential of Table \ref{tab: effective potentials} that behaves in this form for the allowed values of its physical parameters.\footnote{For the Klein-Gordon field the effective potential (\ref{e: effective potential Klein Gordon D4}) is similar to that for the same field propagating in two-dimensional de Sitter spacetime \cite{LopezOrtega:2011sc}. The difference is that the factor multiplying to $(1-r^2)$ is $(\mu^2 -2)$ in four dimensions and $\mu^2$ in two dimensions. Notice that in two-dimensional de Sitter spacetime the effective potential of the massive Klein-Gordon field is always positive, in contrast to the behavior of the effective potential (\ref{e: effective potential Klein Gordon D4}) (see Fig.\ \ref{fig: Klein Gordon D 4 q 0}). }

For the effective potentials that behave in this way we must discuss whether the boundary condition $U(0)=0$ can be imposed, and we believe that for these examples we need a separate analysis. In what follows we do not analyze these potentials, because we like to show some results about the de Sitter QNM that satisfy the boundary conditions (a) and (b) enumerated above.

Establishing a bound on the imaginary parts of the QNF, Horowitz and Hubeny prove that the QNM of the Schwarzschild anti-de Sitter black hole are stable \cite{Horowitz:1999jd}.  Although the exact values of the de Sitter QNF are known, and using these we can discuss the classical stability of the QNM, it is convenient to have an independent proof of their stability and prove it with a different method. Hence we modify the method by Horowitz and Hubeny \cite{Horowitz:1999jd} to show the classical stability of the de Sitter QNM.

Taking the time dependence as $\textrm{exp}(-i \omega t)$ and following Horowitz and Hubeny we propose that the function $U$ of Eq.\ (\ref{e: Schrodinger equation}) satisfies the de Sitter QNM boundary condition near the horizon, that is, this function takes the form 
\begin{equation}
 U = \textrm{e}^{i \omega r_*} \Phi ,
\end{equation} 
where the function $\Phi$ is a solution of the differential equation 
\begin{equation} \label{e: equation Phi}
 f \frac{\dd^2 \Phi}{\dd r^2} +\left( \frac{\dd f}{\dd r} + 2 i \omega \right) \frac{\dd \Phi}{\dd r} - \frac{V}{f} \Phi = 0 ,
\end{equation} 
and it behaves as $\Phi(0)=0$ and near the horizon $\Phi$ goes to a constant.

As in Ref.\ \cite{Horowitz:1999jd} we multiply Eq.\ (\ref{e: equation Phi}) by the complex conjugate of $\Phi$.\footnote{In what follows $\Phi^*$ denotes the complex conjugate of $\Phi$.} Integrating the result we obtain
\begin{equation} \label{e: first integration Phi}
 \int_0^1 \dd r \left[ \Phi^* f \frac{\dd^2 \Phi}{\dd r^2} + \left( \frac{\dd f}{\dd r} + 2 i \omega \right) \Phi^* \frac{\dd \Phi}{\dd r} - \frac{V}{f} \left| \Phi \right|^2 \right]= 0 .
\end{equation} 
Next we integrate by parts the first term of the expression (\ref{e: first integration Phi}) and observe that the surface terms vanish. Thus we get that the previous formula becomes
\begin{equation} \label{e: result first integration Phi}
 \int_0^1 \dd r \left[ \left| \frac{\dd \Phi}{\dd r} \right|^2  f  - 2 i \omega \Phi^* \frac{\dd \Phi}{\dd r} + \frac{V}{f}  \left|\Phi \right|^2 \right] = 0 .
\end{equation} 
We take the imaginary part of  the previous expression \cite{Horowitz:1999jd} to get
\begin{equation} \label{e: imaginary part Phi}
 \int_0^1 \dd r \, \Phi^* \frac{\dd \Phi}{\dd r} = - \frac{\omega^* \left| \Phi(1) \right|^2  }{ \omega - \omega^*},
\end{equation} 
and from the formulas (\ref{e: result first integration Phi}) and (\ref{e: imaginary part Phi}) we find
\begin{equation} \label{e: positive potential condition}
 \int_0^1 \dd r \left[  f \left| \frac{\dd \Phi}{\dd r} \right|^2  + \frac{V}{f}  \left|\Phi \right|^2  \right] =  2 i \omega \int_0^1 \dd r \, \Phi^* \frac{\dd \Phi}{\dd r} = - \frac{ \left| \omega \right|^2  \left| \Phi(1) \right|^2 }{\im (\omega)} .
\end{equation}

Hence if $V \geq 0$ then $\im (\omega) < 0$ and the de Sitter QNM are stable. Thus for the fields of Table \ref{tab: effective potentials} we study whether their effective potentials are nonnegative to prove the classical stability of their de Sitter QNM. For the gravitational and electromagnetic perturbations it is possible to show that for the allowed values of the parameters, their effective potentials are always positive (see Table \ref{tab: effective potentials}). Thus for the gravitational and electromagnetic fields we assert that their de Sitter QNM are stable.

\begin{figure}[tbp]
\begin{center}
\includegraphics[clip]{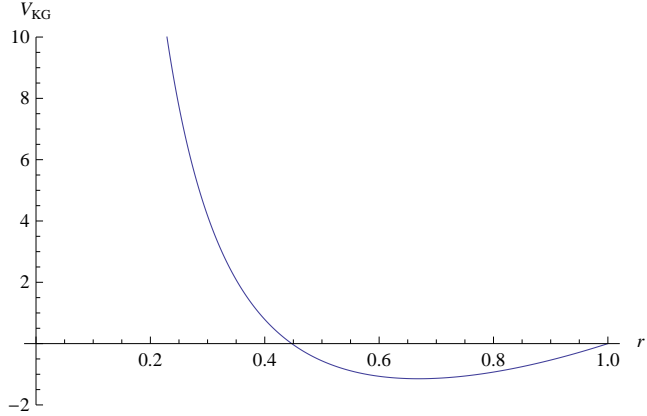}
\caption{For the five-dimensional de Sitter spacetime we draw the effective potential $V_{KG}$ of the Klein-Gordon field with $q=0$ and $\mu = 0$.}
\label{fig: Klein-Gordon D 5 q 0 m 0}
\end{center}
\end{figure}

\begin{figure}[tbp]
\begin{center}
\includegraphics[clip]{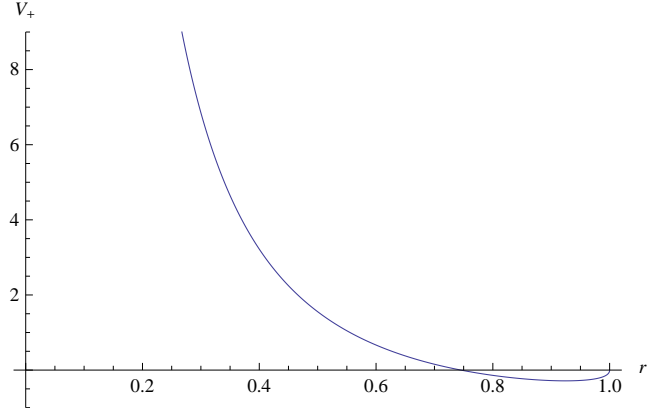}
\caption{For the five-dimensional de Sitter spacetime we plot the effective potential $V_+$ of the massless Dirac field with $\lambda=3/2$.}
\label{fig: Dirac D 5 lambda 3-2 m 0}
\end{center}
\end{figure}

For the Klein-Gordon and Dirac fields propagating in de Sitter spacetime their effective potentials are not strictly positive. For example, for the massless Klein-Gordon field with $q=0$, when $D > 4$, if $1 > r > \sqrt{\frac{D-4}{D}}$ then $V_{KG} < 0$. Moreover, for the massless Dirac field, if $\lambda > 1$, $1 > r^2 > 1 - \frac{1}{\lambda^2}$ then $V_+ < 0$.\footnote{In four-dimensional de Sitter spacetime,  for the Klein-Gordon with $q=0$ and $\mu^2 < 2$, we find $V_{KG} < 0$ in $r \in (0,1)$ (see Fig.\ \ref{fig: Klein Gordon D 4 q 0}) and for the massless Dirac field with $\lambda = 1$ we get that $V_+ < 0$ for $r \in (0,1)$  (see Fig.\ \ref{fig: Dirac vplus lambda 1}). Nevertheless note that these effective potentials do not diverge at $r=0$. }  Notice that for the massless Klein-Gordon field only for the value $q=0$ the effective potential is not strictly positive, whereas for the massless Dirac field the effective potential $V_+$ is not strictly positive for the allowed values of $\lambda$. (See Figs.\ \ref{fig: Klein-Gordon D 5 q 0 m 0}  and \ref{fig: Dirac D 5 lambda 3-2 m 0} for some examples.)


From the formula (\ref{e: positive potential condition}), for these values of the parameters for the Klein-Gordon and Dirac fields we can not assert that the corresponding de Sitter QNM are stable. Hence we need to modify the previous method. We follow to Ishibashi and Kodama \cite{Ishibashi:2003ap} to find
\begin{equation} \label{e: transformation potential}
 \int_0^1 \dd r \left[ f \left| \frac{\dd \Phi}{\dd r} \right|^2  + \frac{V}{f}  \left|\Phi \right|^2  \right] =  \int_0^\infty \dd r_* \left[ \left| \frac{\dd \Phi}{\dd r_*} \right|^2 + V  \left| \Phi \right|^2 \right] = \int_0^\infty \dd r_* \Phi^* \left[ -\frac{\dd^2}{\dd r_*^2} + V  \right] \Phi .
\end{equation} 
We notice that in the third step the boundary terms cancel. In this formula and in what follows $r_*$ is the tortoise coordinate of the de Sitter spacetime.

Defining the differential operator \cite{Ishibashi:2003ap}
\begin{equation}
 \tilde{D} = \frac{\dd }{\dd r_*} + S ,
\end{equation} 
where $S$ is a function of $r$ such that $S(1)=0$, we get from the expression (\ref{e: transformation potential}) 
\begin{eqnarray} \label{e: Phi condition second}
  \int_0^\infty \dd r_* \Phi^* \left[ -\frac{\dd^2}{\dd r_*^2} + V  \right] \Phi &=& \int_0^\infty \dd r_* \left[ (\tilde{D} \Phi)^* (\tilde{D} \Phi)  + \Phi^* \left( V + \frac{\dd S}{\dd r_*} - S^2 \right) \Phi \right] \nonumber \\ 
&=& \int_0^\infty \dd r_* \left[ (\tilde{D} \Phi)^* (\tilde{D} \Phi)  + \Phi^*  \tilde{V} \Phi \right] ,
\end{eqnarray} 
with $\tilde{V}$ given by
\begin{equation}
 \tilde{V} = V + \frac{\dd S}{\dd r_*} - S^2 .
\end{equation} 
Thus for a field such that its effective potential $V$ is not strictly positive, if we find a new potential $ \tilde{V}$ such that $\tilde{V} \geq 0$, then we show the stability of the corresponding de Sitter QNM.

For example, for the Klein-Gordon field we choose \cite{Ishibashi:2003ap}
\begin{equation} \label{e: S function Klein Gordon}
 S = - \frac{(D-2) f}{2 r} ,
\end{equation} 
to obtain
\begin{equation} \label{e: Klein Gordon effective potential tilde}
 \tilde{V} = f \left( \frac{q(D + q -3)}{r^2} + \mu^2 \right) .
\end{equation} 
Since for the allowed values of the physical parameters in de Sitter spacetime this new potential satisfies   $\tilde{V} \geq 0 $, we find that for the Klein-Gordon field its de Sitter QNM are stable. Notice that this result is valid for the effective potentials of the massive and massless Klein-Gordon fields.

For the effective potential $V_+$ of the massless Dirac field we choose
\begin{equation} \label{e: S function Dirac}
 S = - \frac{\lambda f^{1/2}}{r},
\end{equation} 
to get that the new potential is equal to\footnote{Writing the effective potentials of the massless Dirac field as \cite{Chandrasekhar book}
\begin{equation*} 
 V_\pm = W^ 2 \pm \frac{\dd W}{\dd r_*}, 
\end{equation*} 
where $W = \lambda f^{1/2}/ r$, we obtain that the function $S$ is equal to $ S = -W $. }
\begin{equation} \label{e: Dirac effective potential tilde}
 \tilde{V}_+ = 0 .
\end{equation} 
Thus the formula (\ref{e: Phi condition second}) simplifies to
\begin{equation} \label{e: Dirac integral condition}
 \int_0^\infty \dd r_* \left| (\tilde{D} \Phi) \right|^2 > 0,
\end{equation} 
and therefore the de Sitter QNM of the massless Dirac field are stable. 

We note that the functions (\ref{e: S function Klein Gordon}) and (\ref{e: S function Dirac}) satisfy $S(1)=0$. Furthermore the results (\ref{e: Klein Gordon effective potential tilde}) and (\ref{e: Dirac effective potential tilde}) are valid for an arbitrary metric function $f$ that satisfies $f(1)=0$,  $f(0)$ is a finite quantity, and $f$ is positive in $r \in (0,1)$. In particular these results are valid for the de Sitter spacetime.
 
It is convenient to notice that for the Klein-Gordon, Dirac, electromagnetic, and gravitational perturbations our results on the classical stability of their de Sitter QNM are valid for $D \geq 4$. In contrast, it is known that for asymptotically de Sitter spherically symmetric black holes, their classical stability depends on the spacetime dimension in a non trivial way \cite{Ishibashi:2003ap}-\cite{Cardoso:2010rz}. For example, only for $D=4,5,6$ we know analytical proofs of the classical stability of the $D$-dimensional Schwarzschild de Sitter spacetime \cite{Ishibashi:2003ap}, \cite{Ishibashi:2011ws}. For $D=5,\dots,11$ there are numerical calculations that state their classical stability \cite{Konoplya:2007jv}. We do not know analytical or numerical calculations determining the classical stability of Schwarzschild de Sitter black holes for $D \geq 12$.

Furthermore, based on numerical calculations, in \cite{Konoplya:2008au} it is showed that for $D =7,\dots,11$ and for sufficiently large charge and cosmological constant, the Reissner-Nordstr\"om de Sitter black holes are unstable under coupled electromagnetic and gravitational scalar type perturbations. Only for $D=4,5$ we are aware of analytical proofs of the classical stability for Reissner-Nordstr\"om de Sitter black holes \cite{Ishibashi:2003ap}, \cite{Ishibashi:2011ws}. For additional arguments about the instability of these black holes see Ref.\ \cite{Cardoso:2010rz}. Hence the classical stability of the de Sitter QNM is stated in a more general way than the stability of the $D$-dimensional Schwarzschild de Sitter and Reissner-Nordstr\"om de Sitter black holes.

\section{Gravitational QNM of the de Sitter spacetime}
\label{s: gravitational QNM}

As we comment in the previous sections, the QNM of the $D$-dimensional de Sitter spacetime ($D \geq 4$) are determined in \cite{Du:2004jt}-\cite{LopezOrtega:2007sr} and in these references, on their existence we find at least two different results. To settle this problem, based on Refs.\ \cite{Horowitz:1999jd}, \cite{Musiri:2003rv}, \cite{Musiri:2003rs} and with a different method to that of Refs.\ \cite{Natario:2004jd}, \cite{Lopez-Ortega:2006my}, in what follows we calculate the QNF of the three types of gravitational perturbations propagating in $D$-dimensional de Sitter spacetimes. Notice that the method we use in this work is simpler than the procedure of Refs.\ \cite{Natario:2004jd}, \cite{Lopez-Ortega:2006my}, and we hope that it will be useful to solve the mentioned issue. We choose the gravitational perturbations due to its intrinsic physical relevance and since these perturbations are analyzed in \cite{Natario:2004jd}, \cite{Lopez-Ortega:2006my}. 

First we note that in $D$-dimensional de Sitter spacetime the Schr\"odinger type equations (\ref{e: Schrodinger equation}) for the three types of gravitational perturbations simplify to the common form \cite{Natario:2004jd}
\begin{equation} \label{e: gravitational perturbations}
 (1-r^2)\frac{{\rm d}^2 U}{{\rm d} r^2} - 2 r \frac{\dd U}{\dd r} + \left( \frac{\omega^2}{1-r^2} - \frac{B(B+1)}{r^2} + A\right)U = 0,
\end{equation} 
where 
\begin{align} \label{e: A B values gravitational}
B = \frac{2 q + D - 4}{2},  \qquad A = \left\{ \begin{array}{l l} \frac{(D-2)D}{4}   & \,\, \,\,\textrm{tensor type}, \\ \\ \frac{(D-4)(D-2)}{4} & \,\,\,\, \textrm{vector type}, \\ \\ \frac{(D-6)(D-4)}{4} & \,\,\,\, \textrm{scalar type}. \end{array} \right. 
\end{align}

To solve Eq.\ (\ref{e: gravitational perturbations}) we make the change of variable $y = r^2$ \cite{Natario:2004jd}, \cite{Lopez-Ortega:2006my} to obtain 
\begin{equation}
y(1-y) \frac{\dd^2 U}{\dd y^2} + \left( \frac{1}{2} - \frac{3 y}{2} \right)  \frac{\dd U}{\dd y} + \left( \frac{\omega^2}{4(1-y)} - \frac{B(B+1)}{4 y}  + \frac{A}{4} \right) U = 0.
\end{equation} 
Proposing that the function $U$ takes the form 
\begin{equation}
 U = (1-y)^E y^C U_2 ,  
\end{equation} 
we find that if the quantities $E$ and $C$ are solutions of the equations
\begin{equation}
 E^2 + \frac{\omega^2}{4} = 0, \qquad \quad C^2 - \frac{C}{2} - \frac{B(B+1)}{4} = 0, 
\end{equation} 
then the function $U_2$ must be a solution of the differential equation 
\begin{eqnarray} \label{e: equation U2}
 y (1-y) \frac{\dd^2 U_2}{\dd y^2} &+& \left( 2 C + \frac{1}{2} -\left( 2 E + 2 C + \frac{3}{2} \right) y \right) \frac{\dd U_2 }{\dd y} \nonumber \\
&-& \left( C^2 + 2 E C + \frac{C}{2} + E^2 + \frac{E}{2} - \frac{A}{4} \right) U_2 = 0 ,
\end{eqnarray} 
which is a hypergeometric type differential equation \cite{Abramowitz-book}, \cite{Guo-book}
\begin{equation} \label{e: hypergeometric differential}
y(1-y) \frac{{\rm d}^2 U_2}{{\rm d} y^2} + (c - (a +b + 1) y )\frac{{\rm d} U_2}{{\rm d} y} - a b U_2   = 0 .
\end{equation} 

Motivated by the Horowitz-Hubeny method \cite{Horowitz:1999jd}, we choose $E = -i \omega / 2$ to satisfy the boundary condition (a) of the de Sitter QNM near the horizon and $C = (1 + B)/ 2$ to fulfill the boundary condition (b) at $r=0$. Making these choices for $E$ and $C$ we get that the parameters $a$, $b$, and $c$ of Eq.\ (\ref{e: hypergeometric differential}) are equal to
\begin{align}
a &= \frac{B}{2} + \frac{3}{4} - \frac{1}{2}\left( \frac{1}{4} + A \right)^{1/2} - \frac{i \omega}{2}, \quad  \quad c =  B + \frac{3}{2}, \nonumber \\
b &= \frac{B}{2} + \frac{3}{4} + \frac{1}{2}\left( \frac{1}{4} + A \right)^{1/2} - \frac{i \omega}{2}.
\end{align} 

As in Refs.\ \cite{Musiri:2003rv}, \cite{Musiri:2003rs} to obtain the appropriate behavior at the boundaries we impose that the function $U_2$ be a polynomial. Therefore we must satisfy the conditions \cite{Abramowitz-book}, \cite{Guo-book}
\begin{equation} 
  a = - p, \qquad \quad \textrm{or} \qquad  \quad b= -p,  \quad \qquad p=0,1,2,\dots
\end{equation} 
From these equations we get that the QNF of the $D$-dimensional de Sitter spacetime are equal to
\begin{equation} \label{e: quasinormal gravitational}
\omega = -i (q + D - 1 - \eta + 2 p), \qquad \qquad \omega = -i( q + \eta + 2 p),
\end{equation} 
where
\begin{align} \label{e: values q}
\eta = \left\{ \begin{array}{l l} 0,  &\,\,\, \textrm{tensor type}, \\ 1, &\,\,\, \textrm{vector type}, \\ 2, & \,\,\, \textrm{scalar type}. \end{array} \right. 
\end{align}  
We point out that the QNF (\ref{e: quasinormal gravitational}) are identical to those of Refs.\ \cite{Natario:2004jd}, \cite{Lopez-Ortega:2006my} (see for example the formula (37) of \cite{Lopez-Ortega:2006my}). We expect that this method allows us to calculate the QNF of other fields moving in de Sitter spacetime.

Notice that the method used here to calculate the QNF (\ref{e: quasinormal gravitational}) is valid for even and odd $D$, and hence these QNF are valid for all $D$. In contrast to the result that we find in \cite{Natario:2004jd} asserting that the QNF (\ref{e: quasinormal gravitational}) are valid only for $D$ odd. In Ref.\ \cite{Lopez-Ortega:2006my} it is asserted that the QNF (\ref{e: quasinormal gravitational}) are valid for even and odd $D$, but in the procedure of this reference it is necessary to study different solutions for $D$ even  and $D$ odd. Furthermore in the method of Ref.\ \cite{Lopez-Ortega:2006my} the employed solutions and the analysis of the problem are more complicated. We expect that our previous results establish the existence of well defined QNF of the gravitational perturbations in even and odd de Sitter spacetimes. 

Furthermore we note that our conclusion on the existence of well defined de Sitter QNF in even and odd dimensions it is valid for the massless Klein-Gordon field, since in the Schr\"odinger type equations the effective potential for this field is equal to that of the tensor type gravitational perturbations \cite{Natario:2004jd}. In contrast to gravitational perturbations for which $q \geq 2$, for the Klein-Gordon field the angular eigenvalues $q$ take the values $q=0,1,2,\dots$, but this fact does not modify our calculation method or change the result for the de Sitter QNF (\ref{e: quasinormal gravitational}). Hence the de Sitter QNF of the massless Klein-Gordon field are well defined in even and odd dimensions.

Making the change of variable
\begin{equation}
 y = \frac{1+x}{2},
\end{equation} 
we find that for the chosen values of the quantities $C$ and $E$, Eq.\ (\ref{e: equation U2}) becomes
\begin{align}
 (1 - x^2) \frac{\dd U_2}{\dd x^2} &+ \left( B + \frac{1}{2} + i \omega - \left(\frac{3}{2} + B - i \omega +1 \right) x \right) \frac{\dd U_2}{\dd x} \nonumber \\
&- \left( \frac{B(B+1)}{4} + \frac{2(B+1)(1- i \omega)}{4} - \frac{A + \omega^2 + i \omega}{4} \right) U_2 = 0 ,
\end{align} 
where $x \in (-1,1)$. This equation is a  differential equation of Jacobi type \cite{Guo-book}, \cite{Szego-book}
\begin{equation} \label{e: Jacobi differential equation}
 (1-x^2) \frac{\dd^2 U_n}{\dd x^2} +(\beta - \alpha -(\alpha + \beta +2)x ) \frac{\dd U_n}{\dd x} +n(n + \alpha + \beta + 1) U_n = 0,
\end{equation} 
with parameters $\alpha$, $\beta$, and $n$ equal to
\begin{equation} \label{e: parameters Jacobi}
 \alpha = - i \omega, \quad \qquad \beta = B + \frac{1}{2}, \quad \qquad n = \frac{i \omega - B - 3/2 \pm (1/4 + A)^{1/2}}{2} .
\end{equation} 

For $n$ a nonnegative integer, from the last formula of (\ref{e: parameters Jacobi}) we get that the de Sitter QNF (\ref{e: quasinormal gravitational}). Furthermore from the QNF for the three types of gravitational perturbations we obtain that the parameter $\alpha$ satisfies $\alpha  \leq -2$. These values that we find for $\alpha$ have a noticeable consequence.

Let $U_{n_1}$ and $U_{n_2}$ two solutions of Eq.\ (\ref{e: Jacobi differential equation}) for different eigenvalues. Following the usual procedure to show the orthogonality of the eigenfunctions \cite{Guo-book}, we arrive to the formula 
\begin{equation}
 \int_{-1}^{1} (1-x)^\alpha (1+x)^\beta U_{n_1} U_{n_2} \dd x = \frac{(1-x)^{\alpha + 1} (1+x)^{\beta + 1} \left[ U_{n_2} \frac{\dd U_{n_1}}{\dd x} - U_{n_1} \frac{\dd U_{n_2} }{\dd x} \right]^{1}_{-1} }{n_2 (n_2 + \alpha + \beta + 1) - n_1 ( n_1 + \alpha + \beta + 1)} ,
\end{equation} 
which is divergent at $x=1$ for $\alpha < -1$, as for the de Sitter radial functions that we find above.

Hence, the radial functions of the de Sitter QNM are not orthogonal solutions of Eq.\ (\ref{e: Jacobi differential equation}) in the interval $(-1,1)$ with weight $(1-x)^\alpha (1+x)^\beta$ and they can not be the usual Jacobi polynomials which are orthogonal with the previous weight. In Ref.\ \cite{Anninos:2011af} it is said that the de Sitter QNM radial functions involve Jacobi polynomials, without additional comments or discussion. As we say above the radial functions of the de Sitter QNM do not involve the usual Jacobi polynomials \cite{Guo-book}, \cite{Szego-book}. As far as we know this fact is not noted in previous papers.

\section{Summary}
\label{s: discussion}

Based on the work by Horowitz and Hubeny, for the Klein-Gordon, Dirac, electromagnetic, and gravitational perturbations we calculate a bound for the imaginary parts of their de Sitter QNF and using this bound we prove the stability of their de Sitter QNM. Without a doubt, we believe that this alternative proof shows the stability of the de Sitter QNM. Notice that for the Klein-Gordon and Dirac fields we employ a slightly different method that for the electromagnetic and gravitational perturbations.

Furthermore, we remark that for the de Sitter QNM (satisfying the boundary conditions (a) and (b) of Sect.\ \ref{s: stability}) of the Klein-Gordon, Dirac, electromagnetic, and gravitational perturbations, our result on their classical stability is valid for all the spacetime dimensions satisfying  $D \geq 4$. 

Moreover, as a noticeable example, we obtain that for $q=0$ and $\mu^2 = 2$ the Klein-Gordon field propagates freely in the static patch of the four-dimensional de Sitter spacetime (\ref{e: metric de Sitter spacetime}) (see Sect.\ \ref{s: stability} and Appendix). Recently for odd-dimensional de Sitter spacetimes in global coordinates, it is found that several fields propagate freely when they satisfy some conditions \cite{Lagogiannis:2011st}. Our result shows that for $q=0$ and $\mu^2 = 2$ the Klein-Gordon field  behaves in a similar way in the static patch of the four-dimensional de Sitter spacetime.

Using a simple method we calculate the de Sitter QNF of the gravitational perturbations. Our results coincide with those of Ref.\ \cite{Lopez-Ortega:2006my}, but the method of Sect.\ \ref{s: gravitational QNM} is simpler than the procedure of Refs.\ \cite{Natario:2004jd}, \cite{Lopez-Ortega:2006my}. We expect that these calculations establish, in odd and even spacetime dimensions, the existence of well defined de Sitter QNF for massless fields. 

We also find that for the gravitational perturbations, the radial functions of their de Sitter QNM are solutions of the Jacobi differential equation, but these radial functions are not orthogonal in the interval $(-1,1)$ with the usual weight $(1-x)^\alpha (1+x)^\beta$, since for the de Sitter QNF (\ref{e: quasinormal gravitational}) we get that the parameter $\alpha$ satisfies $\alpha \leq -2$. We think that this fact deserves further research.

\section{Acknowledgments}

This work was supported by CONACYT M\'exico, SNI M\'exico, EDI-IPN, COFAA-IPN, and Research Projects SIP-20120773 and SIP-20121648.

\begin{appendix}

\section{On the de Sitter QNF of the Klein-Gordon field with  $q=0$, $\mu^2 = 2$, and  $D=4$}
\label{s: appendix}

For the Klein-Gordon field with $q=0$ and propagating in four-dimensional de Sitter spacetime, from the formula (\ref{e: effective potential Klein Gordon D4}) we observe that if $\mu^2 = 2$ the effective potential of the Schr\"odinger type equation goes to zero. Therefore the solutions of Eq.\ (\ref{e: Schrodinger equation}) take the form (\ref{e: solution near the horizon}) for all $r \in (0,1)$, and hence we can not satisfy the boundary conditions of the de Sitter QNM. Thus for these values of the parameters we find that the formula (51) of Ref.\ \cite{Lopez-Ortega:2006my} does not give valid QNF of the de Sitter spacetime.

To see where the procedure of Ref.\ \cite{Lopez-Ortega:2006my} fails for these values of the physical parameters, we note that for $q=0$ the radial function  regular at $r=0$ is (see the formulas (46) and (48) of \cite{Lopez-Ortega:2006my})
\begin{equation} \label{e: radial function four de Sitter}
 R = C_1 (1-y)^{i \omega / 2} {}_{2}F_{1}(a,b;c;y) ,
\end{equation} 
with 
\begin{equation} \label{e: a b c Klein Gordon}
 a= \frac{i \omega}{2} + 1, \qquad \qquad b = \frac{i \omega}{2}  + \frac{1}{2}, \qquad \qquad c=\frac{3}{2}.
\end{equation} 

We point out that the quantities $a$, $b$, and $c$ of the formulas (\ref{e: a b c Klein Gordon}) satisfy
\begin{equation}
 a - b + 1 = c ,
\end{equation} 
and therefore we write the radial function (\ref{e: radial function four de Sitter}) as (see the formulas (15.4.14) and (8.6.9) of \cite{Abramowitz-book})
\begin{align}
 R &= C_1 \Gamma(3/2) y^{-1/4} (1-y)^{-1/2} P^{-1/2}_{-i \omega / 2 - 1/2} \left( \tfrac{1+y}{1-y} \right) \nonumber \\
&=  - \frac{C_1}{2 i \omega y^{1/2} } \left( (1+y^{1/2})^{- i \omega} (1-y)^{i \omega / 2} - (1+y^{1/2})^{ i \omega} (1-y)^{- i \omega / 2} \right) ,
\end{align} 
where $\Gamma$ denotes the gamma function and $P^a_b$ the associated Legendre functions.

From the last expression we find that near the horizon the radial function (\ref{e: radial function four de Sitter}) contains ingoing and outgoing waves. Notice that we can not cancel the ingoing  part of the radial function without cancel the outgoing part. Thus in four-dimensional de Sitter spacetime, for the Klein-Gordon field with  these values of $q$ and $\mu$, the method of Ref.\ \cite{Lopez-Ortega:2006my} does not work, since we can not satisfy the boundary conditions of the de Sitter QNM.

It is convenient to note that the absence of the de Sitter QNF of the Klein-Gordon field for $q=0$, $\mu^2 = 2$, and $D=4$ is an exceptional situation, and it is caused by the fact that for these values of $q$, $\mu$, and $D$ the effective potential $V$ of the Klein-Gordon field is equal to zero. As far as we can see, only for these values of the physical parameters we expect that the de Sitter QNM do not exist.

\end{appendix}

\end{document}